\documentclass[useAMS,usenaib]{mn2e}
\usepackage{graphicx}
\title[VLT and GTC observations of the type 2 quasar SDSS J0123+00]
{VLT and GTC observations of SDSS J0123+00: a type 2 quasar triggered in a galaxy encounter?\thanks
{Based on observations carried out at the
European Southern Observatory (Paranal, Chile) with FORS2 on VLT-UT1 (programme 
69.A-0123(A)) and at the Observatory Roque de los Muchachos (La Palma, Spain) with OSIRIS on GTC (programme GTC27-09A).}}
\author[Villar-Mart\'\i n et al.]{M. Villar-Mart\'\i n$^{1}$, C. Tadhunter$^{2}$, E. P\'erez$^{1}$, A. Humphrey$^{3}$,   A. Mart\'\i nez-Sansigre$^{4,5}$,
\newauthor R. Gonz\'alez Delgado$^{1}$, M. P\'erez-Torres$^{1}$ \\
$^{1}$Instituto de Astrof\'\i sica de Andaluc\'\i a (CSIC), Glorieta de la Astronom\'\i a s/n, 18008 Granada, Spain. montse@iaa.es\\
$^{2}$Dept. of Physics and Astronomy, University of Sheffield, Sheffield S3 7RH,
 UK\\
$^{3}$Instituto Nacional de Astrof\'\i sica, Optica y Electr\'onica (INAOE), Apt
do. Postal 51 y 216, 72000 Puebla, Mexico\\
$^{4}$Astrophysics, Department of Physics, University of Oxford, Keble Road, Oxford OX1 3RH, UK\\
$^{5}$Institute of Cosmology and Gravitation, Univ. of Portsmouth, Dennis Sciama Building, Burnaby Road, Portsmouth, PO1 3FX, UK}

\begin{document}

\date{Accepted 2010 May 27.  Received 2010 May 25; in original form 2009 September 21}

\pagerange{\pageref{firstpage}--\pageref{lastpage}} \pubyear{2002}

\maketitle

\begin{abstract}

We present  long-slit spectroscopy, continuum and [OIII]$\lambda$5007 imaging data obtained with the Very Large Telescope  and the
Gran Telescopio Canarias  of the   type 2 quasar SDSS J0123+00 at $z=$0.399. The quasar lies in a complex, gas-rich environment. It appears to be physically connected by a tidal bridge to another galaxy at a projected distance of $\sim$100 kpc, which suggests this is an interacting system.  Ionized gas is detected to a distance of at least $\sim$133 kpc from the nucleus. The nebula has a total extension of $\sim$180~kpc. This is one of the largest ionized nebulae ever detected associated with an active galaxy. Based on the environmental properties, we propose that the origin of the nebula is tidal debris from a galactic encounter, which
could as well be the triggering mechanism of the nuclear activity. SDSS J0123+00 demonstrates that 
 giant, luminous ionized nebulae can exist associated with type 2 quasars of low  radio luminosities, contrary to expectations based on type 1 quasar studies.
\end{abstract}


\label{firstpage}

\section{Introduction}

 Previous works suggest that the most
luminous  extended emission line regions (EELRs)  around type 1 quasars (QSOs) at redshift $z\la$0.5 exist preferentially associated with
 steep-spectrum radio-loud QSOs with luminous [OIII]$\lambda$5007 nuclear emission and low broad line region
metallicities ($Z\la$0.6$Z_{\odot}$). These EELRs have low metallicities as well and seem especially prevalent in QSOs showing signs of strong interaction. The origin of these nebulae is controversial, whether cold accretion of intergalactic gas, tidal debris from a galactic merger or remnants of galactic superwinds (for a review see Fu \& Stockton 2009; Stockton, Fu \& Canalizo 2006).

Very little is known about the  existence of  EELRs associated with type 2 quasars (e.g. Humphrey et al. 2009; Gandhi, Fabian \& Crawford 1996). 
These objects 
are unique laboratories to investigate the existence and properties of quasar EELR.  The fortuitous occultation
of the active  galactic nucleus (AGN) acts like a 
``natural coronograph'',
allowing a detailed study of many properties  of the surrounding medium, without the
problems associated with the bright quasar
point spread function. On
 the other hand, due to their much lower radio luminosity,  distortions   
imprinted by the radio activity on the properties of
the underlying nebulae are  less important than in radio-loud
objects.

For these reasons, we are undertaking a project based on imaging, long-slit and integral field
data obtained with FORS2/VLT, the integral field spectrograph PMAS on the 3.5m telescope at Calar Alto Observatory (Humphrey et al. 2009), and the tunable filter
OSIRIS  on the GTC. These are being used  to investigate the existence of EELR associated with Sloan Digital Sky Survey (SDSS) type 2 quasars 
at $z\sim$0.3-0.4 and characterize their morphological,
ionization and kinematic properties.

In this paper we present results  on the type 2 quasar   SDSS
J012341.47+004435.9 (SDSS J0123+00 hereafter)
at $z=$0.399 (Zakamska  et al. 2003).
The object  was  detected by the  NVSS survey with a flux
density of   11.7$\pm$0.6 mJy  (Condon et al. 1998).  The radio
emission is  spatially
 unresolved. Assuming a radio spectral
index of $\alpha=0.7$ (where $L_{\nu}\propto \nu^{-\alpha}$), the
expected radio luminosity density at 5 GHz is 5$\times$10$^{30}$ erg
s$^{-1}$ Hz$^{-1}$ sr$^{-1}$. This value suggests that the quasar is
``radio-quiet" or at most ``radio-intermediate" (Miller, Peacock \& Mead
1990).

We assume
$\Omega_{\Lambda}$=0.7, $\Omega_M$=0.3, H$_0$=71 km s$^{-1}$ Mpc$^{-1}$.
In this Cosmology,  1$\arcsec$ corresponds to 5.33 kpc at $z$=0.399.

\section[]{Observations and analysis}

The VLT observations were carried out on the 8th and 9th Sept 2008 using FORS2 on UT1.
 The 691$\_$55+69 filter was used to obtain the intermediate band image on the 1st night. It has a spectral window 6635-7185 \AA\ which covers both continuum and
 redshifted H$\beta$ and [OIII]$\lambda\lambda$4959,5007. The pixel scale is 0.25 arcsec pix$^{-1}$.
A diamond pattern mosaic of 4 images   with individual exposures of 200 seconds was obtained, shifting each
image by 10$\arcsec$ in RA or Dec relative
to the previous one. This was done  to  minimize the impact of bad pixels and ghosts on the CCD. 
The images were combined, with simultaneous cosmic ray removal after bias subtraction.  The seeing 
size measured from the final image is 
FWHM=0.95$\pm$0.05$\arcsec$ for the 1st night.

Long slit spectroscopy  was obtained at position angles (PA) PA60 (first night) and PA76 (second night), aligned with  extended diffuse emission
or nearby potentially companion objects revealed by the images (Fig.~1).
The 600RI+19 grism and the   GG435+81 order sorting filter were used with
 a 1$\arcsec$ wide slit. The useful spectral range was $\sim$5030-8250 \AA\, corresponding to rest frame 3595-5901 \AA. The object was moved along the slit between
two different positions 
for a more accurate sky subtraction.
 The data were reduced following standard procedures
with STARLINK and IRAF packages.
The spectra were debiased, flat fielded, wavelength ($\lambda$) calibrated, background subtracted, combined with simultaneous cosmic ray removal, flux calibrated and corrected for Galactic extinction.
Geometric distortion was found to be negligible.

The nights were photometric. Comparison of several spectrophotometric standard stars taken with a 5$\arcsec$ slit during the run gave a  flux calibration
accuracy of 5\% over the entire spectral range. 
To perform the spectroscopic analysis, the line profiles were fitted with 
Gaussian functions.
The FWHM values were corrected for instrumental broadening (IP=7.2$\pm$0.2 \AA\, as measured from the sky emission lines).   Point sources did not fill the slit  during the second night (seeing FWHM=0.65$\pm$0.05$\arcsec$)
and this introduces uncertainties on the kinematic analysis.  For this reason,  the PA60 spectrum was
used to characterize the  kinematic properties at the different
spatial regions identified in the spectra. 

GTC images were obtained with the OSIRIS camera on 17th Oct 2009
using the order sorting filter 694/44. The $2\times2$ detector binning
provided a spatial scale of 0.254 arcsec pix$^{-1}$.  A continuum image and a continuum+[OIII]$\lambda$5007 image were produced. 
Each one consisted of three 200 second exposures  obtained through the tunable filters 
with FWHM=16 \AA\ and  central $\lambda$ ($\lambda_C$)  at the optical axis  of 7019 \AA\ and 7069 \AA\ respectively.
Due to the nature of the instrument $\lambda_C$ changes with radial
distance from the optical axis.  $\lambda_C$ values 
 were 7002.9 and 7052.7 \AA\ for the on-band and off-band images respectively at the location of the target, corresponding to rest-frame [OIII]$\lambda$5007 and the adjacent continuum. The average seeing was 0.8$\pm$0.1$\arcsec$.
The observations were reduced and flux calibrated with the standard star G158-100 using IRAF routines. 
Tunable filter profiles were generated at  $\lambda_C$ corresponding to the spatial location of G158-100 on the CCD 
and applied to its absolute flux spectrum to obtain the flux calibration factors.
The individual target exposures for each filter were sky subtracted, spatially aligned
 and combined.

\section{Results}

\subsection{Imaging data}

The VLT intermediate band image is shown in Fig.~1 (top panel). The bottom panel shows the GTC pure [OIII]5007 (green) and continuum  (red) composite image.
The original [OIII]+continuum image was continuum subtracted, and therefore the green colour shows the pure gas emission. 
The central elliptical galaxy presents 
 distorted contours due to a detached ``blob'' which is more evident in the GTC images and emits both continuum and strong line emission.
  Low surface brightness diffuse features and  knots are  detected  well outside the optical size of 
 the galaxy.  A long filament extends roughly in the  E-W direction, to a maximum projected distance from the nuclear region of
 $r\sim$14$\arcsec$ (75 kpc) and a total  extension of 25$\arcsec$ (133 kpc) 
(throughout the text distances and sizes are projected values).

 The filament is clearly detected in [OIII] extending towards and overlapping with galaxy ``G1'', which emits not only continuum but also line emission  since it is  clearly detected in the pure [OIII] GTC image. Therefore it is at similar $z$ as 
the quasar. Both objects seem to be physically connected by the filament mentioned above, which is reminiscent of a tidal bridge.
   ``G2'', which  is detected both in
continuum and   [OIII], also lies at a similar $z$ and  is located roughly along the direction of the filament. Other galaxies are found  near  the quasar in projection, but we cannot confirm that any of them are at similar $z$.
 
The VLT and GTC images show that SDSS J0123+00 lies in a rich gaseous environment and suggest that it is a member of an interacting system.

\begin{figure}
\includegraphics{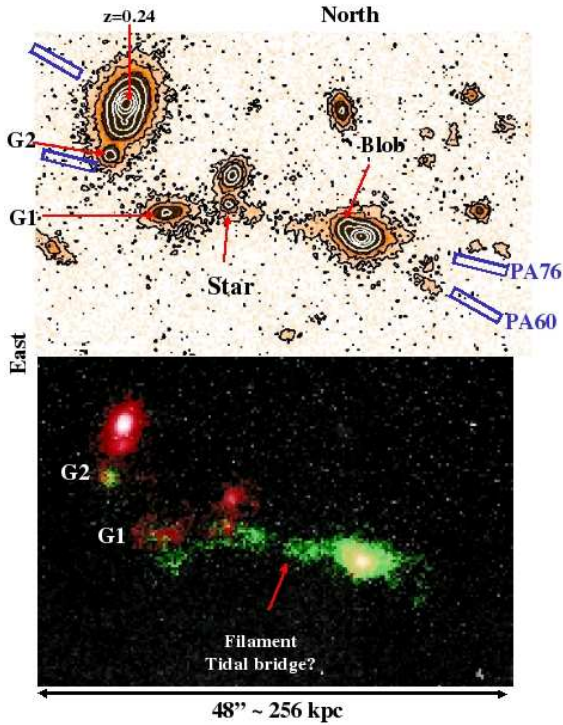}
\vspace{3.8in}
\caption{Top: FORS2-VLT intermediate band image of SDSS J0123+00 containing continuum, H$\beta$ and [OIII]$\lambda\lambda$4959,5007 emission. The field size is 48$\arcsec\times$31.5$\arcsec$ or 254 kpc $\times$ 168 kpc. 
 The contours (arbitrary units) 
 highlight the low surface brightness features.
 The slit positions used for the  VLT spectra  are
indicated. Bottom: GTC continuum
subtracted [OIII]5007 image (green) and continuum image
(red), color coded in the region of interest.
  The images demonstrate that 
``G1'' and ``G2'' are at the same $z$ as the QSO. The  gaseous filament is reminiscent of a tidal bridge connecting the  ``G1''galaxy 
and the quasar. The images suggest that SDSS J0123+00 is a member of an interacting system. }
\end{figure}

\subsection{Long slit spectra.}

The 2-dim spectra at PA60 and PA76 including the [OIII]$\lambda\lambda$4959,5007 doublet
are presented in Fig.~2. The low surface brightness  features revealed
by the images  (Fig.~1) along the slit direction emit a rich emission
line spectrum. Continuum
emission is not detected from these features.
The spectra reveal a total extension 
of the ionized nebula of 12.5$\arcsec$  (67 kpc) and $\sim$34$\arcsec$ (180 kpc) 
 along  PA60 and PA76 respectively. The maximum radial extent from the nucleus is $r\sim$25$\arcsec$ (133 kpc)
measured along PA76 towards the East. The filament revealed by the images  is part of this structure.

The visual inspection of the 2-dim spectra 
reveal 3 distinct spatial regions where the gas is characterized by different kinematic and ionization
properties, which are
 recognisable at both PA. We have named them 
the central region, the intermediate region and the  low surface brightness
nebula (LSBN hereafter). We have analysed them separately. Since it is possible that within each region
there is a spatial variation of the kinematic and ionization properties, we have defined
within each region one or more apertures (Ap. hereafter) as specified below.

In order to characterize the ionization properties 
of the gas at each spatial location, we plot the line ratios for the different apertures 
in the [OIII]/H$\beta$ vs. [OII]$\lambda$3727/[OIII]$\lambda$5007  diagnostic diagram (Fig.~3), which
involves the strongest emission lines 
detected in most apertures. Only apertures for which at least
two emission lines are detected are shown.
 For comparison, the standard photoionization
model
sequence often applied to low redshift type 2 active galaxies is also shown  (see Villar-Mart\'\i n et al. 2008 for a discussion of this
model sequence applied  to type 2 quasars). The models assume solar metallicity, gas density $n$=100 cm$^{-3}$ and  power-law index $\alpha$=-1.5.  The ionization parameter $U$ varies along the sequence.
A subsample of  HII galaxies extracted
from the catalogue by Terlevich et al. (1991) is also shown.

\begin{figure}
\includegraphics{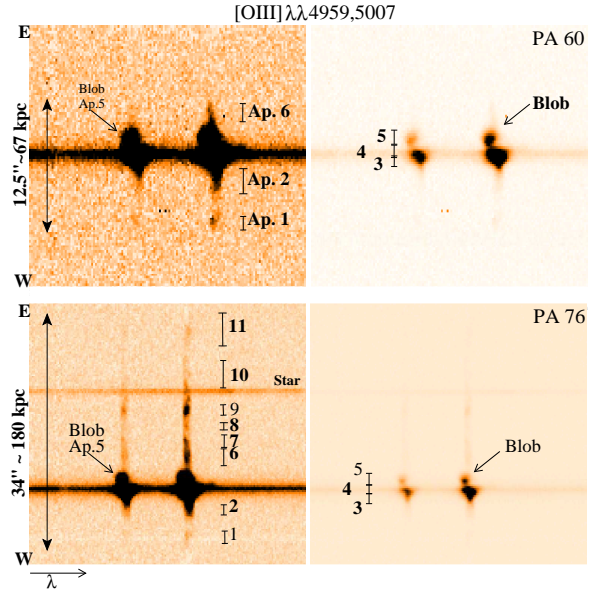}
\vspace{3.0in}
\caption{2-dim FORS2-VLT spectra of the [OIII] doublet at PA60 (top) and PA76 (bottom). The spectra
are shown in two different color scales to highlight the low surface brightness regions (left) or
the regions closer to the nucleus (right). The apertures
used for the spectroscopic analysis (see text) are shown with numbers. 
Notice the striking extension $\sim$180 kpc of the knotty, extended nebula along PA76. The extension to the East encompasses
the filament in Fig.~1.}
\end{figure}

\begin{itemize}

\item {\it $r \la$ 1$\arcsec$. The central region (Ap. 4 in Fig.~2 and 3)}. This is the region in which
the continuum emission is brightest.
  The continuum
spatial profile is strongly peaked and is dominated by an unresolved 
central component (FWHM=0.92$\pm$0.03$\arcsec$, PA60), although diffuse continuum emission is also detected in excess of the PSF wings up to a total
extension of $\sim$4$\arcsec$ or $\sim$21 kpc. The central region emits 
a  rich emission line spectrum, typical of AGN (Fig.~3 dark blue solid circle), including strong HeII$\lambda$4686.  The FWHM([OIII]) value measured from the 1-dim spectrum is 430$\pm$30 km s$^{-1}$.

We have fitted the continuum in Ap. 4 stellar population models. The nebular continuum was first removed.
  Its contribution, 6\% of the total continuum flux at 4020 \AA\ ($z=$0), was estimated from the H$\beta$ flux. A reasonable range of parameters (gas metallicities,
temperature and density) and no reddening (as implied by the H$\beta$/H$\gamma$=0.48$\pm$0.07 value) were considered.  
 The fit (Fig.~4) to
the  nebular continuum subtracted spectrum  implies
the presence of an  old stellar population     and an excess of blue light which cannot be accounted for with  only old stars.  If it is due to young stars, the best fit is obtained 
with an old stellar population of $>$5-10 Gyr of age that contributes with $\sim$80\% of the continuum at 4020 \AA\ ($z=$0) 
 plus a young
 stellar population (age$\leq$ 10 Myr) which contributes  $\sim$20\%. 
 However, based on these data alone we cannot distinguish
 if this blue excess  is alternatively due  to scattered nuclear radiation (the polarization level measured by Zakamska et al. (2006) was 
(2.7$\pm$2.2\%).

\begin{figure}
\includegraphics{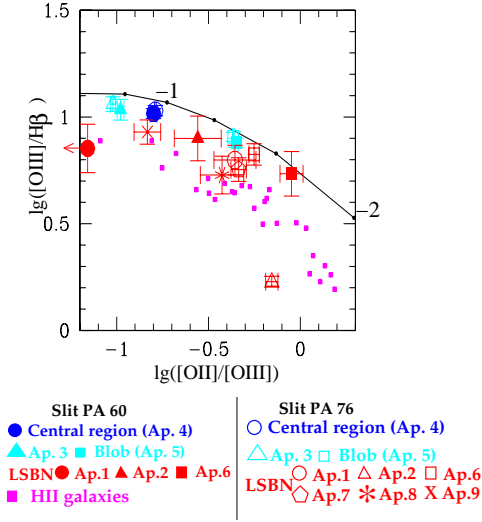}
\vspace{2.8in}
\caption{Diagnostic diagram and the gas ionization properties. Different symbols are used for
different apertures. PA60: Solid symbols. PA76: Hollow symbols. Blue and cyan symbols correspond
to the central and intermediate regions apertures, while red symbols correspond to apertures
across the LSBN. The small magenta symbols are HII galaxies
from Terlevich et al. (1992) catalogue.  The solid black line
represents standard AGN photoionization models (see text). Numbers -1 and -2 indicate the log(U) values corresponding to
the nearest models.}
\end{figure}

\begin{figure}
\includegraphics{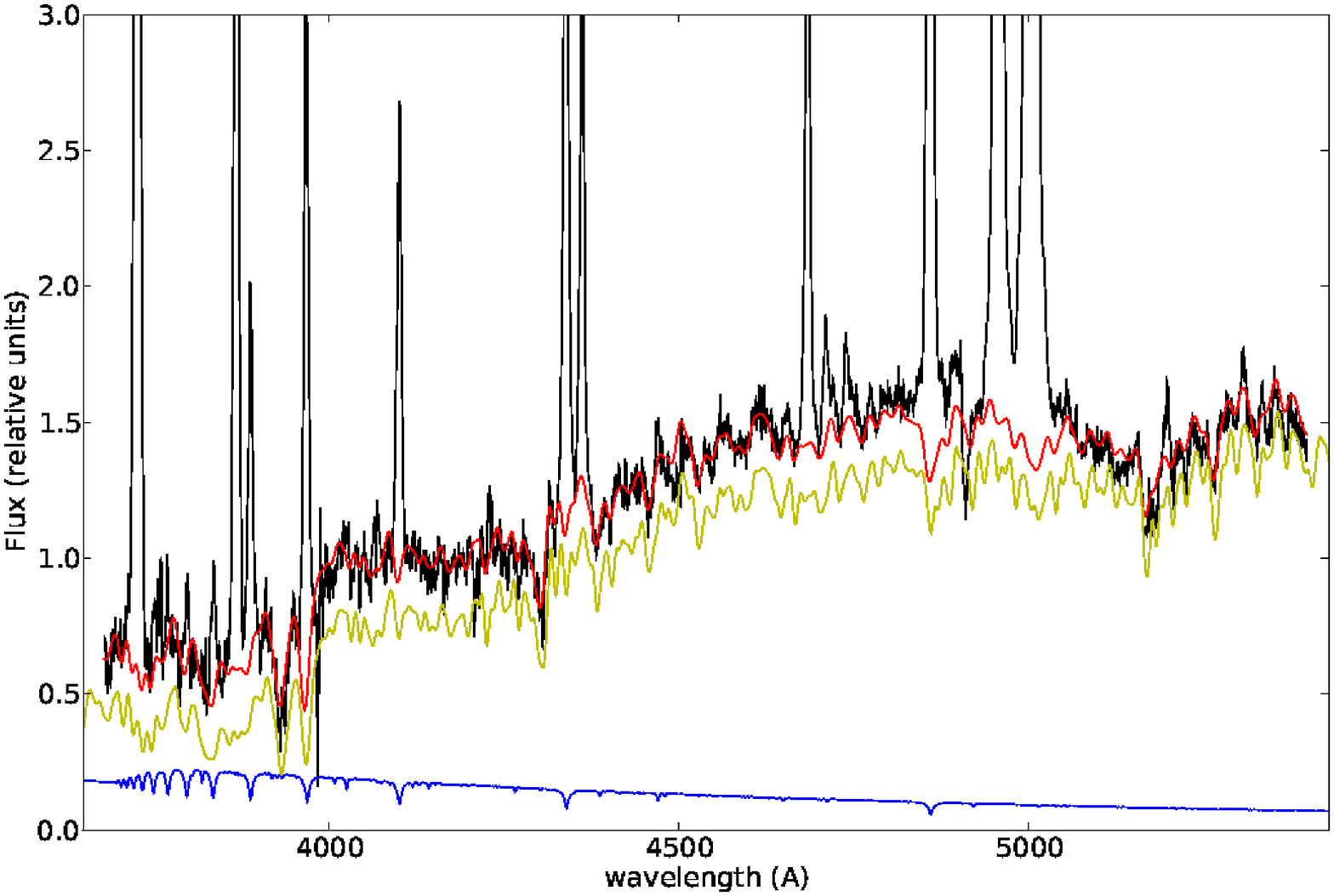}
\vspace{1.7in}
\caption{Nuclear spectrum (black) (the nebular continuum has been
removed) and the best fit stellar population model (red) 
obtained with the code STARLIGHT (Cid Fernandes et al. 2005).
An excess of blue light  above the old stellar population (yellow)  is implied by the analysis. We do not know whether this excess is
due to young stars (blue)  and/or scattered AGN light. }
\end{figure}

\item {\it  1 $\arcsec < r \la$ 2.5$\arcsec$. The intermediate region (Ap.~3 and 5 Fig.~2 and 3)}.   It
extends at both sides of the central region,  up to a distance of 2.0$\arcsec$ to the E  and $\sim$2.6$\arcsec$ to the W 
  respectively from the continuum
centroid (PA60). 
   The location of Ap.~3 and 5 in the diagnostic diagram (Fig.~3, cyan solid square and triangle)
shows that they are predominantly photoionized by the AGN.
 This is also confirmed by
the presence of strong HeII emission. 

An interesting feature within the intermediate  region, which is clearly seen
in the GTC images (Fig.~1)  is a detached blob, which
looks  rather compact (Ap. 5, Fig.~2). FWHM values along PA60 (1.15$\pm$0.05$\arcsec$)  and  PA76 (1.1$\pm$0.1$\arcsec$) implying a 
physical size of a few kpc.
Its centroid is located at $\sim$1.3$\arcsec$ from the
continuum centroid. Faint continuum emission is  detected.
The lines are blueshifted relative to the rest of the ionized
gas (Fig.~2)  and show very asymmetric line profiles which reveal complex kinematics.

\item {\it $r>$3$\arcsec$. The low surface brightness nebula (LSBN)} has a total extension of  $\sim$180 kpc,  as mentioned above (Fig.~2), and
overlaps towards the East with the filament or tidal bridge identified in the 
images (Fig.~1). The fact that the FIRST and NVSS fluxes are consistent with each other shows that no significant radio
emission at 1.4 GHz is present on scales $\ga$5$\arcsec$ or 27 kpc and therefore any jet present is probably much smaller than the nebula. The spectrum integrated across Ap. 6 to 9 along PA76 is shown in Fig.~5.

\begin{figure}
\includegraphics{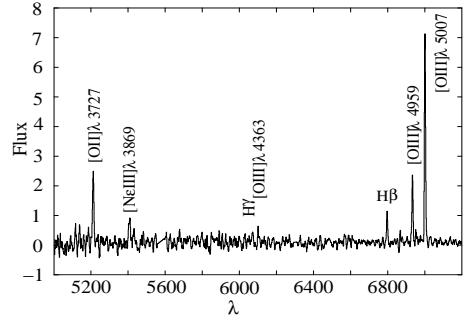}
\vspace{1.5in}
\caption{LSBN spectrum integrated from Ap. 6 to 9 along PA76. Flux in units of 10$^{-17}$ erg s$^{-1}$ cm$^{-2}$ \AA$^{-1}$.}
\end{figure}

A study of the emission line kinematics of the LSBN along PA60 shows that they are quiescent: it emits very narrow, unresolved emission lines 
with FWHM$\la$100 to $\la$140 km s$^{-1}$ depending on spatial location, and the rest-frame velocity shifts relative to the nucleus are
also small ($\Delta V \la$200 km $s^{-1}$). The lines are narrower than the IP
measured from the sky lines at most spatial locations along PA76, confirming the general narrowness of the lines also in  this direction.

In order to investigate the spatial variation of the ionization properties of the LSBN, several apertures where considered along the slit. 
Since the LSBN is knotty and inhomogeneous, we selected apertures that isolate individual features when possible (Fig.~2).  We have plotted the
 location of the apertures along PA60 and PA76 in the diagnostic diagram (Fig.~3). 
It is important to note that the standard AGN photoionization models,
although successful at reproducing many spectral properties 
of type 2 active galaxies
in general and  type 2 quasars in particular,
cannot explain the large scatter in the diagnostic diagrams presented by
this class of objects (e.g. Robinson et al. 1987), with a substantial fraction
 overlapping with  HII galaxies.
The locations of the LSBN apertures of SDSS J0123+00
in the diagrams are consistent with the scatter presented by the general
population of type 2 active galaxies. Although a range
 of gas and/or AGN continuum properties has usually been claimed as responsible for the scatter, 
 Villar-Mart\'\i n et al. (2008) proposed that an alternative possibility
is that  young stars contribute  to the
gas excitation, in addition to the AGN. We explore this possibility further for SDSS J0123+00.

Fig. 3 shows that   Ap. 2 and 6 line ratios along PA60, which are closer to the AGN, are
consistent with pure AGN photoionization, while moving outwards (Ap. 1, at $r\sim$6.6$\arcsec$ or 35 kpc),  the line ratios are consistent with HII star forming galaxies. 
A similar trend is found along PA76: Ap.~6
presents an intermediate spectrum, 
 with line ratios  between
the standard AGN models and HII galaxies (HeII is also detected). 
All other apertures ($r\ga$4.5$\arcsec$ or 24 kpc, most clearly  Ap.  2, 7, 8, 9) 
overlap with the   HII galaxy zone. 

Therefore, although with the available information it is not possible to  discard that the entire nebula is
photoionized by the AGN, we  find that the spatial variation of the LSBN line ratios
can be explained if 
 the inner regions are photoionized by the AGN,
while moving outwards stellar photoionization becomes dominant. Line ratios based on [OI]$\lambda$6300, H$\alpha$, [NII]]$\lambda$6583, [SII]$\lambda\lambda$6716,6732  would help to discriminate between the two
scenarios.

 In the stellar photoionization scenario,
 we can estimate the gas oxygen abundance in regions purely ionized by stars, using  the R23 method (Pilyugin \& Thuan 2005).   The spectrum
integrated across the LSBN has  H$\gamma$/H$\beta$=0.43$\pm$0.07,
consistent within the errors with  the
 theoretical case B recombination value, and thus we assume no line reddening.
 To make sure that AGN photoionization is relatively negligible, we have
used the spectra of those apertures along  PA76 (e.g. Ap. 7,8,9) which  are  unambiguously placed among the HII galaxies 
and far from the standard AGN models in Fig.~3, taking the errorbars into account. 

The individual spectra, as well as the integrated spectra across all these apertures, are too noisy 
 to set useful constraints on the [OIII]$\lambda$4363 flux and $T_e$. This introduces  
 uncertainties on the abundance
 determination,
due to the bimodal shape of the 12+log(O/H) vs. log(R23) function (Pilyugin \& Thuan 2005).  The $R23$ log values measured  
for Ap. 7, 8 an 9 along PA76 are 1.01$\pm$0.06, 0.96$\pm$0.09 and 1.10$\pm$0.06 respectively and imply
abundances well below solar, in the range 8\% to 42\% solar taking errors and uncertainties into account,
and depending on the aperture. Similarly  abundances are often found in compact
HII galaxies (e.g. Guzm\'an et al. 1997, Terlevich et al. 1991).
For comparison, we have used the SDSS spectrum to constrain the abundance within the inner  AGN photoionized region using the
method discussed in Humphrey et al. (2008). The resulting oxygen abundance  must be $\ga$25\% solar in order to reproduce the optical
line ratios.

The narrowness of the spectral lines compared with the IP (see \S2),
  implies that the spatial
extent in the direction perpendicular to the slit is $<$1$\arcsec$ or $<$5.3 kpc. For the brightest knots  isolated along the slit, 
 FWHM$\sim$1.1$\pm$0.1$\arcsec$   corresponding to 4.7$\pm$0.6 kpc, taking the seeing  
into account.  Sizes $\la$5 kpc are also measured for the UV rest frame continuum knots identified in the HST images of Zakamska et al. (2006).

 The  H$\beta$ luminosity for Ap. 7,8,9
 is in the range (0.4-1.0)$\times$10$^{40}$ erg s$^{-1}$. If the gas is photoionized by stars,   
star forming rates (SFR) $\sim$0.09-0.23 M$_{\odot}$  yr$^{-1}$ are implied
(Kennicutt 1998; given the large EW of the lines,
stellar  absorption of H$\beta$ is expected to be negligible).   Considering the small knot sizes,
slit flux losses could be $\la$30\%. This uncertainty does not
affect our conclusions. Similar SFRs and sizes are found for some nearby compact HII galaxies 
(Telles, Mu\~noz-Tu\~n\'on, Tenorio-Tagle 2001, Guzm\'an et al. 1997).

\end{itemize}

 \section{Discussion and conclusions}

 SDSS0123+00 at $z=$0.399 is associated with a giant, luminous ionized nebula, which extends for  $\ga$180 kpc.  
This is one of the largest EELRs ever detected associated with an AGN.  The EELR $L_{[OIII]}\sim$3$\times$10$^{42}$ erg  s$^{-1}$ is similar to that
of very luminous EELRs associated with type 1 quasars. Therefore, this object demonstrates
that   very luminous and giant EELR can also be associated with low radio luminosity quasars, contrary to expectations based on type 1 quasar studies (see \S1). 
 The low EELR abundances inferred ($<$0.5$Z_{\odot}$) and the clear signs of interactions  are features often found  in type 1 quasars with
luminous EELR. 

SDSS0123+00 lies in a gas  rich environment and  it seems to be a member of an interacting system. The quasar appears to be physically connected
by  a tidal bridge  to a companion galaxy at $\sim$100 kpc in projection. The giant EELR encompasses the  bridge.
Although we cannot discard the possibility that the entire EELR is photoionized by the AGN,  the spatial variation of the line ratios can be explained if the gas in the inner regions ($r\la$24 kpc) is predominantly ionized by the AGN while stellar photoionization
becomes dominant as we move outwards. The star formation might have been triggered during the interaction.  Whatever ionizes the gas, we find that the  properties 
(sizes, line widths, line ratios and luminosities) of the individual knots isolated within the LSBN are consistent with nearby HII galaxies.

Based on the environmental properties  (the tidal bridge and the companion galaxy ``G1''), we propose that 
the  origin of the   EELR in this object is tidal debris from a galactic encounter. The companion  must contain a substantial amount of
metal poor gas, to explain the subsolar
EELR abundances.
The nuclear activity  might have
been triggered in such encounter, rather than in the final stages of a major galaxy merger (as the nuclei coalesce). Probably the closest approach in the encounter occurred $>$100 Myr ago, assuming
a typical 300 km s$^{-1}$ separation speed, and a true distance between the two galaxies of 100 kpc. Such a scenario would be consistent with recent imaging results for powerful radio galaxies (Ramos Almeida et et. 2010, in prep.).

\section*{Acknowledgments}
Thanks to an anonymous referee for useful comments. Work funded with support from the Spanish MICINN (grants AYA2004-02703,
AYA2007-64712 and AYA2009-13036-C02-01).
 AMS is supported by a UK STFC postdoctoral fellowship.  EP is grateful to Ignacio Gonz\'alez Serrano for useful discussions and help with OSIRIS data. Thanks to the GTC and VLT staff for their support during the observations.

\end{document}